\input epswor.sty
\documentstyle[12pt]{article}

\def\PRL#1{{\it Phys.~Rev.~Lett.~}{\bf #1}}
\def\PRD#1{{\it Phys.~Rev.~}{\bf D#1}}

\def\NPB#1{{\it Nucl.~Phys.~}{\bf B#1}}

\def\PLB#1{{\it Phys.~Lett.~}{\bf B#1}}

\def\ZPhysC#1{{\it Z.~Phys.~}{\bf C#1}}

\def\ProgTP#1{{\it Prog.~Th.~Phys.~}{\bf #1}}

\def\beq{\begin{equation}}
\def\eeq{\end{equation}}
\def\beqa{\begin{eqnarray}}
\def\eeqa{\end{eqnarray}}

\def\Slash{\hskip -.6em/}

\def\la{~\mbox{\raisebox{-.6ex}{$\stackrel{<}{\sim}$}}~}

\def\Slash#1{%
   \setbox0=\hbox{$#1$}\dimen0=\wd0\setbox1=\hbox{/}\dimen1=\wd1%
   \ifdim\dimen0>\dimen1\rlap{\hbox to \dimen0{\hfil/\hfil}}#1%
   \else\rlap{\hbox to \dimen1{\hfil$#1$\hfil}}/\fi}%

\begin{document}

\rightline{SLAC-PUB-7127}
\medskip
\rightline{hep-ph/9603253}
\medskip
\rightline{March 1996}
\bigskip
\bigskip
\renewcommand{\thefootnote}{\fnsymbol{footnote}}

{\centerline{\bf A NEW SUPERSYMMETRIC CP VIOLATING CONTRIBUTION}}
{\centerline{{\bf TO NEUTRAL MESON MIXING}}}
\footnotetext{Research supported
by the Department of Energy under contract DE-AC03-76SF00515}

\bigskip
{\centerline{Mihir P. Worah}}
\smallskip
\centerline {\it Stanford Linear Accelerator Center}
\centerline {\it Stanford University, Stanford, CA 94309}

\bigskip

{\centerline{\bf Abstract}

We study the contribution to flavor changing neutral current 
processes from box diagrams with light higgsinos and squarks. 
Starting with just the Cabbibo Kobayashi Maskawa (CKM) phase, 
we find contributions to the $K^0$
and $B^0$ meson mass matrices that are out of phase with the Standard
Model contributions 
in the case of substantial mixing between the up-type squarks.
This difference in phase could be large enough to
be detected at the proposed $B$ factories, with interesting
implications for the unitarity triangle of CKM matrix elements. 

\vspace{2.0in}

%{\centerline{Submitted to Physical Review D.}}

\newpage

\section{Introduction}
\renewcommand{\thefootnote}{\arabic{footnote}}
\setcounter{footnote}{0}

Flavor changing neutral currents (FCNC) are predicted by, and
constrain most extensions of the Standard Model. 
They will be well studies at the proposed $B$ factories both in the
hope of uncovering new physics, as well as pinning down the parameters
of the Standard Model more accurately. In this regard it is important
to see what the predictions of various extensions of the Standard
Model are. The minimal supersymmetric standard model (MSSM) is a well
motivated and popular extension of the Standard Model \cite{DefMssm}.
Its predictions
for FCNC have been extensively studied both in the
constrained version \cite{Dugan, Masiero1, Hagelin, Branco}, and for
the more general case \cite{Masiero2, Nir1}
(the first of Refs.
\cite{Masiero2} also considers the consequences of embedding the MSSM
in various GUTs).

We partially redo the analysis of the previous papers, concentrating,
however, on a region of supersymmetric parameter space that has not
been explored in this context: namely the case of low $\tan\beta$ 
($\tan\beta$ denotes the ratio of vacuum expectation values of
the two Higgs bosons in the MSSM),
where the lightest charginos are mostly higgsinos, and where the
right-handed up-type squarks are light and substantially mixed with
each other.{\footnote{This particular scenario, in the absence of
squark mixing, has been recently studied in detail in \cite{Zwirner}. 
The absence of squark mixing, however, precludes
the existence of the interesting CP violating effects we discuss
here.}}

Although this scenario cannot occur in the constrained MSSM, our
motivation for studying it are twofold. Firstly, it has
recently been proposed that light higgsinos and right-handed stops in
the small $\tan \beta$ limit
could partially explain the anomolously large $Z \rightarrow b\bar b$
partial width \cite{Finnel, Wells}.{\footnote{Mixing between the
squarks reduces the effect on the $Z
\rightarrow b\bar b$ partial width, but allieviates the problems the
scenario of \cite{Wells} has with the decays of the top quark.}} As we
will explain in the next
section, this particular configuration of the theory
leads to robust predictions for supersymmetric
contributions to neutral meson mixing.
Secondly, even if one starts with just the Cabbibo Kobayashi Maskawa
(CKM) phase, mixing between the right-handed up-type squarks 
can give rise to large
contributions to the neutral meson mixing matrices that are out of
phase with the usual Standard Model contributions.
Thus, for example, the angle $\beta'_{KM}$ measured by the CP
asymmetry in the decay $B_d \rightarrow \Psi K_S$ is not the angle
$\beta_{KM}$ of the CKM matrix and the unitarity triangle constructed
using the angle $\beta'_{KM}$ will fail to close. It is the
feasibility of uncovering this interesting possibility that we wish to
explore. 

In section~2 we show how these large FCNC can arise for light
higgsinos and right handed squarks, and derive formulas for the new
contributions to the $K-\bar K$, $B_d-\bar B_d$ and $B_s-\bar B_s$
mixing marices. We study the implications of these new contributions
to the angles and lengths of the unitarity triangle, and the
experimental tests of this scenario in Section~3. Section~4 contains
our conclusions.

\section{Light Higgsinos and FCNC}

The importance of light higgsinos and squarks to FCNC processes is
simply that the presence of the quark masses at the
quark-squark-higgsino vertices removes the super GIM cancellation
present in the usual supersymmetric box graphs. Thus higgsino mediated
box graphs give non-zero contributions to neutral meson mixing
independent of details of the squark mass matrices. 
Given the factor of $m_t/\sin\beta$ 
that appears in the $d_L - \tilde{u}_R - \tilde{h}$ vertex, we are
gauranteed to get large supersymmetric contributions to $K-\bar K$, 
$B_d-\bar B_d$ and $B_s-\bar B_s$ mixing for light right-handed
up-type squarks, higgsinos, and small $\tan\beta$.{\footnote{A similar
argument could be made for charged Higgs boson mediated boxes. These,
however, are generally smaller because of the heavy top quark in the
loops. We have checked that over the parameter range we study here,
the charged Higgs bosons can be made heavy enough to satisfy
constraints from other processes like $b \rightarrow s \gamma$, and
not affect our analysis. }}

Consider the basic quark-squark-higgsino vertex:
\beq
{\cal L}_I = \frac{g}{\sqrt 2 M_W \sin\beta}
             \bar d_L[V^{\dagger}_{KM}\hat M_U \tilde V_U]
             \tilde u_R \tilde h
\label{vertex}
\eeq
where $V_{KM}$ is the CKM matrix, $\hat M_U$ is the diagonal matrix of
up-type quark masses, and $\tilde V_U = V^{\dagger}_{UR}\tilde V_{UR}$
is a product of the unitary matrices that diagonalize the right-handed
up-type quark and squark mass matrices respectively. Starting with
Eq.~(\ref{vertex}) we can derive the following very simple formula for
the supersymmetric contribution to the off diagonal terms in the 
mass matrix of the neutral
meson consisting of the quarks $(\bar a b)$ with $a, ~b=1,2,3$:
\beq
(M_{ab})_{12} = K_{\bar a b}[V^{\dagger}_{KM}\hat M_U \tilde V_U 
               \frac{{\tilde M}^{-1}}{2\sqrt 3\sin^2\beta}
               {\tilde V}^{\dagger}_U\hat M_U V_{KM}]_{ab}^2
\label{master}
\eeq
where
\beq
K_{\bar a b} = \frac{G_F^2}{12\pi^2}(B_{\bar a b}
                f_{\bar a b}^2m_{\bar a b}\eta)
\label{defK}
\eeq
with $B_{\bar a b},~f_{\bar a b},~m_{\bar a b}$ being 
the bag factor, decay constant and mass of the meson,
and $\eta$ a QCD correction factor which we always set equal to the
corresponding QCD correction for the Standard Model box diagram with
top quarks in the loop. 
$\tilde M$ is the diagonal matrix of right-handed up-type squark
masses. We have ignored any difference between the charged higgsino
mass and the squark masses in deriving Eq.~(\ref{master}). 
This approximation does not significantly
affect the accuracy of our results for the range of masses we
consider (this was noted in \cite{Zwirner}), while allowing
us to derive the simple expression of Eq.~(\ref{master}).

Let us now assume the following form
for the mixing matrix $\tilde V$:
\beq
\tilde V = \left(\begin{array}{ccc}
            1& 0& 0\\
            0& 1/\sqrt 2& 1/\sqrt 2 \\
            0& -1/\sqrt 2& 1/\sqrt 2 \end{array} \right)
\eeq
{\it i.e.} the right-handed scalar charm and top are maximally mixed.
This form of the mixing matrix could be 
motivated in some of the models for fermion masses based on Abelian
horizontal symmetries ~\cite{Nir1, Nir2}. 
We will denote the common mass of the lightest squark and charged
higgsino by $\tilde m$, and assume no special degeneracy between the
physical squark masses
In this case, we 
can use Eq.~(\ref{master}) to obtain the
following expressions for the meson mixing to first order in $m_c/m_t$:
\beq
M_{12}^{B_d} ={\displaystyle{
              \frac{G_F^2}{12\pi^2}B_{B_d}f_{B_d}^2m_{B_d}\eta_{B_d}m_t^2
               [a V_{td}^{\ast 2}V_{tb}^2}}
               +{\displaystyle{b V_{td}^{\ast 2}V_{tb}^2 +
                 c V_{cd}^{\ast}V_{td}^{\ast}V_{tb}^2]}}
\label{massbd}
\eeq
\beq
M_{12}^{B_s} ={\displaystyle{
              \frac{G_F^2}{12\pi^2}B_{B_s}f_{B_s}^2m_{B_s}\eta_{B_s}m_t^2
               [a V_{ts}^{\ast 2}V_{tb}^2}}
                +{\displaystyle{b V_{ts}^{\ast 2}V_{tb}^2 +
                c V_{cs}^{\ast}V_{ts}^{\ast}V_{tb}^2]}}
\label{massbs}
\eeq
\beqa
M_{12}^{K} =&{\displaystyle{
            \frac{G_F^2}{12\pi^2}B_{K}f_{K}^2m_K\eta_{K}m_t^2
            [a V_{td}^{\ast 2}V_{ts}^2}}
             +{\displaystyle{b V_{td}^{\ast 2}V_{ts}^2 +
               c  V_{cd}^{\ast}V_{td}^{\ast}V_{ts}^2
           + c V_{td}^{\ast 2}V_{cs}V_{ts}}} \nonumber \\
            &+{\displaystyle{\frac{m_c^2\eta_{cc}}{m_t^2\eta_{K}}
             f_2(y_c) V_{cd}^{\ast 2}V_{cs}^2 +
             \frac{m_c^2\eta_{ct}}{m_t^2\eta_{K}}
             f_3(y_c,y_t)
              V_{cd}^{\ast}V_{cs}V_{td}^{\ast}V_{ts}}}]
\label{massk}
\eeqa
where
\beq
a=f_2(y_t),~
b=\frac{1}{48\sin^4\beta}\frac{m_t^2}{{\tilde m}^2},~
c=\frac{1}{24\sin^4\beta}\frac{m_cm_t}{{\tilde m}^2},
\label{defabc}
\eeq
$y_i=m_i^2/M_W^2$,
and the functions $f_2(x),~f_3(x,y)$ are defined in ~\cite{Inami,
London1}:
\beqa
f_2(x)&=&{\displaystyle{\frac{1}{4}+\frac{9}{4(1-x)}-
         \frac{3}{2(1-x)^2}-\frac{3x^2\ln x}{2(1-x)^3}}}\nonumber \\
f_3(x,y)&=&{\displaystyle{\ln(\frac{y}{x})-\frac{3y}{4(1-y)}
           (1+\frac{y\ln y}{1-y})}}
\eeqa

The terms proportional to $b$ and $c$ in
Eqs.~(\ref{massbd}-\ref{massk}) are the
supersymmetric contributions, and have important consequences for the
determination of the CKM matrix elements as we show in the next section. 
The dominant supersymmetric 
contribution proportional to $b$ is present also in the usual analyses
based on the constrained MSSM, and is always in phase with the 
Standard Model contribution proportional to $m_t^2$. Although we
started with only the CKM phase, in this model, 
both the $B_d-\bar B_d$ and the $K-\bar K$
mass matrices have a second out of phase contribution given by the
term proportional to $c$. This
contribution is a result of the mixing between the right-handed scalar
top and charm, and should be observable at the $B$ factory $CP$
violating experiments. An estimate of the importance of this term
compared to the ``in phase'' supersymmetric contribution is given by
\beq
\frac{c|V_{cd}|}{b|V_{td}|} 
\simeq \frac{2m_c}{A\lambda^2 m_t}
\simeq 50 \frac{m_c}{m_t}. 
\eeq
where $A$ and $\lambda$ parametrize elements of the CKM matrix as
shown below, and we have used $(A\lambda^2)^{-1}=|V_{cb}|^{-1}\simeq
25$. 

\section{The Unitarity Triangle}

In the Wolfenstein parametrization \cite{Wolfenstein}, the CKM matrix
is given by
\beq
V_{CKM}=\left(\begin{array}{ccc}
        1-\frac{1}{2}\lambda^2 & \lambda & A\lambda^3(\rho-i\eta)\\
         -\lambda & 1-\frac{1}{2}\lambda^2 & A\lambda^2 \\
         A\lambda^3(1-\rho-i\eta) & -A\lambda^2 & 1 \end{array}\right)
\eeq
We can best visualize the effects of the new contributions
on the CKM parameters by plotting the allowed regions in
the $\rho-\eta$ plane.{\footnote{The parameter $\lambda$
corresponds to the Cabbibo angle,
and is extremely well measured in tree-level standard model
decays. Although the parameter $A$ occurs in all of the expressions
for neutral meson mixing, and we allow it to vary in our subsequent
fits, its best fit value is always close to that determined from the
CKM element $V_{cb}$ whose determination is again dominated by
tree-level standard model physics. Thus the effects of new physics are
dominantly felt by the parameters $\rho$ and $\eta$.}} 
We will plot the usual three constraints
coming from the experimentally measured quantities
$|V_{ub}|/|V_{cb}|, ~\Delta m_{B_d}$, and $|\epsilon|$, 
as well as the constraint from $Arg(M^{B_d}_{12})$ which will be 
cleanly measured at the $B$ factories by the $CP$ asymmetry in the 
decay $B_d \rightarrow \Psi K_S$.
Although the model satisfies the constraint
from $\Delta m_K$, we do not include it in the subsequent analysis
because of the large uncertainty in the standard
model prediction for this quantity due to long distance effects.
 
The curves to be plotted are determined by the following equations
\beq
1)~~~~~~~~~~~~~~~~~~
\frac{|V_{ub}|}{|V_{cb}|}=\lambda \sqrt{\rho^2+\eta^2}
\label{expt1}
\eeq
this determines a circle centered at the origin of the $\rho-\eta$
plane. Since this quantity is determined by tree-level decays, it is
not affected by the presence of new physics.
\beq
2)~~~~~~~~~~~~~~~~~~
\Delta m_{Bd} = 2|M_{12}^{B_d}|
\eeq
which gives
\beq
(1-\frac{c}{2A\lambda^2(a+b)}-\rho)^2+\eta^2=
\frac{\Delta m_{B_d}}{2 K_{\bar b d} m_t^2 (a+b)}
\label{expt2}
\eeq
where $K_{\bar b d}$ and $a,~b,~c$
have been defined in Eqs.~(\ref{defK},\ref{defabc}). This once again
determines a circle on the $\rho-\eta$ plane. The presence of new
physics has two effects here. The in phase supersymmetric contribution
given by $b$ in the denominator on the right-hand-side of
Eq.~(\ref{expt2}) reduces the radius of the circle, and the out of
phase contribution proportional to $c$ displaces the center from $\rho
= 1$.
\beq
3)~~~~~~~~~~~~~~~~~~~
|\epsilon| = \frac{Im M_{12}^{K}}{\sqrt 2 \Delta m_K}
\eeq
which gives
\beq
\eta[(a+b)A^2\lambda^4(1-\rho)-c A\lambda^2(1-\rho)-\frac{c}{2}
A\lambda^2 + P'_0]=\frac{\Delta m_K |\epsilon|}{\sqrt 2 K_{\bar s
d} m_t^2 A^2\lambda^6}
\label{expt3}
\eeq
with
\beq
P'_0=\frac{m_c^2}{m_t^2}[\frac{\eta_{ct}}{\eta_{K}}f_3(y_c,y_t)-
                            \frac{\eta_{cc}}{\eta_{K}}]
\eeq
This curve determines a hyperbola in the $\rho-\eta$ plane, with the
new physics once again having two effects. The term proportional to
$b$ reduces the distance of the directrix from the origin, 
and the term proportional to $c$ shifts the coordinates to which the
hyperbola is referred.
Finally, 
\beq
4)~~~~~~~~~~~~~~~~~~~~~~~~~
V_{td}=|V_{td}|e^{-i\beta_{KM}}
\eeq
which leads to the straight line
\beq
\eta=(1-\rho) \tan \beta_{KM}
\eeq
where $\beta_{KM}$ is determined from the expression
\beq
Arg(M_{12}^{B_d})=\tan^{-1}[\frac{(a+b)\sin 2\beta_{KM}-c'\sin\beta_{KM}}
                                 {(a+b)\cos 2\beta_{KM}-c'\cos\beta_{KM}}],
\label{expt4}
\eeq
with
\beq
c'=\frac{c}{A\lambda^2\sqrt{(1-\rho)^2+\eta^2}}.
\eeq
Here the shift from the standard model expectation is entirely due to
the ``out-of-phase'' contribution proportional to $c$, which 
tends to increase the phase of $B_d-\bar
B_d$ mixing as compared to the standard model 
(in Eqs.~(\ref{expt2},~\ref{expt3},~\ref{expt4}), the standard
model limit can be recovered by setting $b = c = 0$. This corresponds
to the limit $\tilde m \rightarrow \infty$).

We plot the constraints from these curves 
using the inputs from Table 1 in
Figs. 1. Our inputs are the same as those in \cite{London2} except
for $|V_{cb}|$ where we use the value in \cite{Zwirner}, and
that we have been slightly less conservative in our estimates of the
uncertainties in $\sqrt {B_B}f_B$ and $B_K$.
\begin{table}
\begin{center}
\begin{tabular}{|l|l|}   
\hline
Parameter & Value \\
\hline
$|V_{ub}|/|V_{cb}|$ & $0.08 \pm 0.02$ \\
$\Delta m_{B_d}$ &  $(0.306 \pm 0.0158) \times 10^{-12}$ GeV \\
$|\epsilon|$ & $ (2.26 \pm 0.02) \times 10^{-3}$ \\
$|V_{cb}|$ & $0.039 \pm 0.002$ \\
$\lambda$ & 0.2205 \\
$m_t$ & $170 \pm 10$ GeV \\
$m_c$ & $1.3$ GeV \\
$\sqrt{B_B}f_B$  & $180 \pm 30 MeV$ \\
$B_K$ & $0.8 \pm 0.2$ \\
$\eta_B$ & 0.55 \\
$\eta_K, ~\eta_{ct}, ~\eta_{cc}$ & 0.57, 0.47, 1.32 \\
$\tilde m, ~\tan\beta$ & 85 GeV, 1 \\ 
\hline
\end{tabular}
\end{center}
\caption {Input parameters for the $\rho-\eta$ analysis
presented in Fig. 1 and Table 2.}
\end{table}
\begin{figure}
\epsfig{file=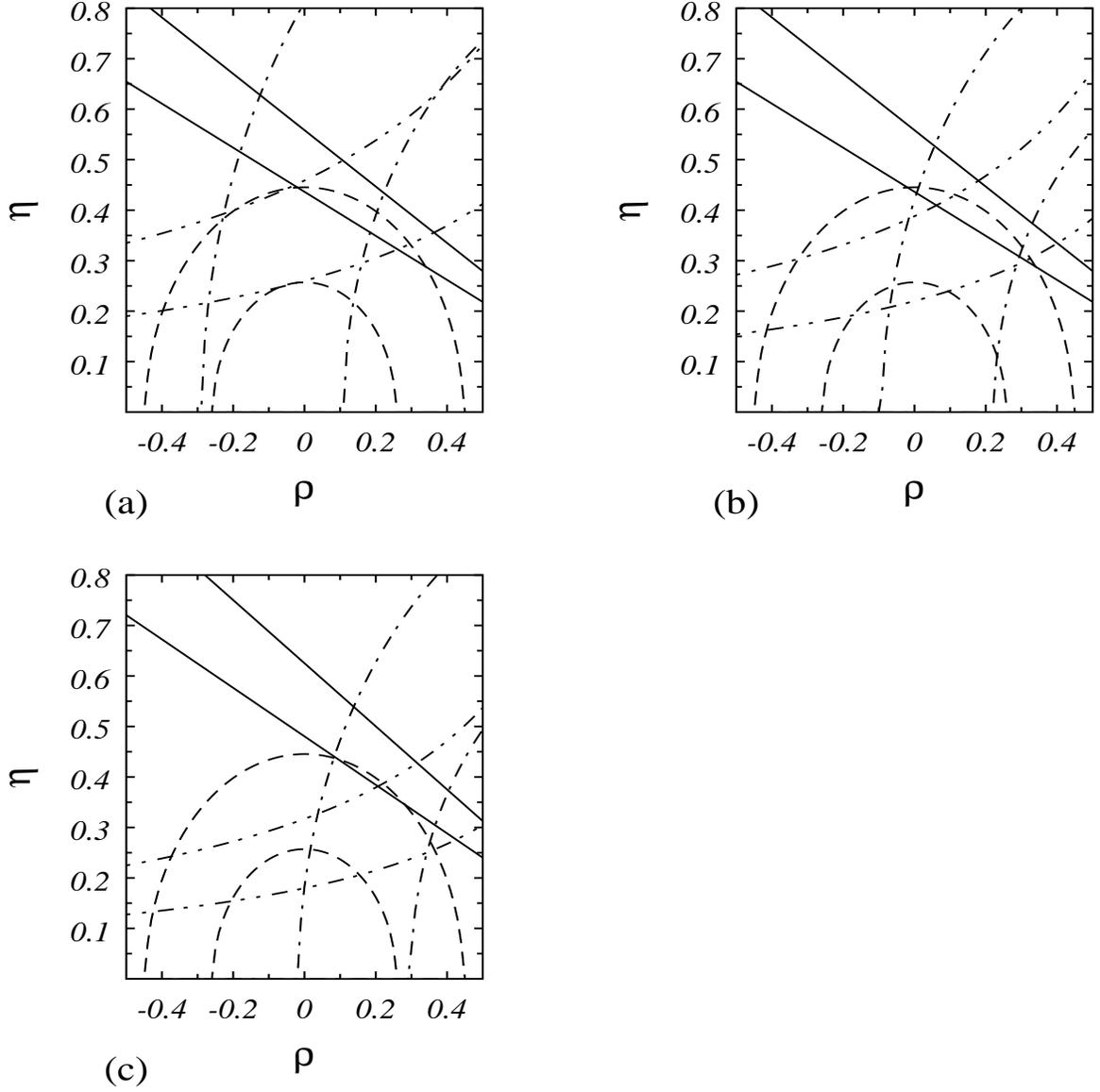,height=6in,width=6in}
\caption {Constraints on $\rho$ and $\eta$ based on the
parameters of Table 1.
The small circles (dash) are from
$|V_{ub}|/|V_{cb}|$, the large circles (dot dash) are from
$\Delta m_{B_d}$, the hyperbolae (dot dot dash) from $\epsilon$ and the
straight lines (solid) from $Arg(M_{12}^{B_d})$.
(a) The standard model. (b) The correct supersymmetric analysis. (c)
The incorrect supersymmetric analysis.}
\end{figure}
Fig. 1(a) corresponds to the standard model case {\it i.e.}
no new physics. Fig.1(b) and Fig. 1(c) include 
the supersymmetric contribution
with $\tilde m=85$ GeV, and $\tan\beta=1$. However Fig. 1(c) contains
the wrong analysis where we incorrectly assume that the supersymmetric
contribution is always in phase with the standard model one. 
In all of these figures we include the error from $m_t$ only in its
effect on the leading coefficients of Eqs.~(\ref{massbd},~\ref{massk}),
ignoring its effect on the terms in the square brackets. We also
ignore the effects of the error in $|V_{cb}|$.
The straight lines in the figures
corresponding to $Arg(M_{12}^{B_d})$ are obtained in the
following way: we determine the point 
%(marked by the cross on the plots) 
in the overlap region of the
other three curves that give us the largest values for
the phase $\beta_{KM}$. This is then plugged into Eq.~(\ref{expt4}),
and we include an error of $\pm 0.059$ in the determination of 
$\sin(Arg(M_{12}^{B_d}))$ as quoted in \cite{Babar}.

Comparing Fig. 1(a) with Figs. 1(b) and 1(c), we notice
that although there
is a large overlap between the allowed regions for the Standard Model
and for the supersymmetric case, it could be possible that 
the supersymmetric contributions, as discussed above shift, $\rho$ and
$\eta$ into a region excluded by the Standard Model. This
possibility which has been noticed in Refs.~\cite{Branco, Zwirner} 
is not very interesting from the point of view of the 
$B$-factory CP violation
experiments. This is because supersymmetric particles in the mass
range we are cosidering would already be detected at the high energy
colliders like LEP2 before the $B$-factories turn on, thus we would know
already that the Standard Model analysis is incorrect. More
interesting is the fact that the line denoting $Arg(M^{B_d}_{12})$
lies outside the allowed region for some values of the phase
$\beta_{KM}$, a possibility that can never occur in the constrained
version of the MSSM. 
Thus, if we incorrectly interpreted this phase as measured by the CP
asymmetry in $B_d \rightarrow \Psi K_S$ as the CKM phase $\beta_{KM}$
(as in Fig. 1(c)),
we would come to the false conclusion that the unitarity triangle does
not close, and that there
are additional sources of CP violation in the theory besides the
complex Yukawa couplings between the quark and Higgs fields.

The obvious way to check for this supposed deviation from unitarity 
is by the clean measurement of phases
that the CP violation experiments at the B-factories allow. 
The CP violating rate asymmetries in the decays 
$B_d \rightarrow \pi\pi$,
$B_d \rightarrow \Psi K_S$, $B_s \rightarrow \rho K_S$ measure the
quantities $\sin2\alpha'$, $\sin2\beta'$, and $\sin2\gamma'$ where 
\beq
2\alpha'=Arg(M^{B_d}_{12})+2\gamma,~~
2\beta'=Arg(M^{B_d}_{12}),~~
2\gamma'=Arg(M^{B_s}_{12})+2\gamma.
\eeq
and $\gamma$ is the phase of $V_{ub}$. In the standard model,
$Arg(M^{B_d}_{12})=2\beta_{KM}$ and $Arg(M^{B_s}_{12})=0$ so the
measured quantities
reduce to -$\sin2\alpha$, $\sin2\beta_{KM}$, and $\sin2\gamma$ (after
making the replacement $\beta_{KM}+\gamma = \pi - \alpha$) where 
$\alpha$, $\beta_{KM}$ and $\gamma$ are the angles of the ``unitarity'' 
triangle. 
Thus if the phases of $B_d-\bar B_d$ or $B_s-
\bar B_s$ mixing are affected by new physics, the three measured 
angles $\alpha'$, $\beta'$ and $\gamma'$ will not correspond to the
angles of the unitarity triangle, and
in general will not add up to $180^{\circ}$. 
This test, however, does not work in
this case, because although the phase of $B_d-\bar B_d$ mixing is
affected by the supersymmetric contribution, 
that of $B_s -\bar B_s$ mixing isn't [Eq. (\ref{massbs})]. 
Thus, if we repeat the
above analysis making the replacement $\beta'+\gamma=\pi-\alpha'$, we
will still obtain a triangle that closes, with angles
$\alpha-\delta,~\beta_{KM}+\delta$ and $\gamma$, where $\delta$ is the
amount by which the phase of $B_d-\bar B_d$ mixing is shifted from the
standard model value. 
This possibility that the angles measured by the above CP
violating experiments would still add up to $180^{\circ}$ if the phase
of $B_d-\bar B_d$ mixing is changed, but that of $B_s -\bar B_s$
mixing isn't was pointed out in ~\cite{NirSilver}. 
An alternative method to measure $\gamma$ is in the $CP$ violating
decay $B_d \rightarrow D^0_{CP}K^{\ast}$ \cite{Dunietz}. 
However, since this measurement is a result of interfering tree-level
amplitudes, it is not affected by the new physics, and we would
still measure the true angle $\gamma$. 
Thus, as in the case above, we would mistakenly
interpret the three angles obtained as summing to $180^{\circ}$.

Another interesting manifestation of this new phase in the 
$B_d-\bar B_d$ mixing matrix could be in the existence of $CP$
violating asymmetries in decays where the standard model predicts
none. A simple example of this is the penguin mediated decay $B_d
\rightarrow K_SK_S$ where the phase of the top mediated penguin
exactly cancels the phase $-2\beta_{KM}$ of $B_d-\bar B_d$ mixing 
in the
standard model. In the model we are considering, this cancellation
would not be exact because of the new phase in $B_d-\bar B_d$ mixing
matrix, and there could be observable $CP$ asymmetries in the
decay. It has recently been observed, however, that sub-dominant
penguins mediated by up and charm quarks could contribute to $CP$
violation in this channel \cite{Fleisher}. We have checked that this
contribution is not only comparable in magnitude to the one due to the
new mixing phase, but is also uncertain in sign. Thus the observation
(or non observation) of $CP$ violation in this decay could not
distinuish this model from either the standard model or the
constrained MSSM.  

The considerations of the previous paragraphs show us that only phase
information is not enough to tell us that we are wrong in assuming that
supersymmetric contributions do not modify the phase of neutral $B$ meson
mixing. In order to detect this, we need to combine the phase
information from the $CP$ violating experiments with 
independent information on
magnitudes (and phases) of the CKM matrix elements available in the
quantities $|V_{ub}|/|V_{cb}|$, $\Delta M_{B_d}$ and $|\epsilon|$
discussed earlier.
To this end we do a $\chi^2$ analysis for the central values of $\rho$
and $\eta$ using the quantities listed in Table 1. Our experimental
inputs are $|V_{ub}|/|V_{cb}|$, $\Delta M_{B_d}$, $|\epsilon|$,
$|V_{cb}|$, $m_t$, and ``projected values'' for $\sin 2\beta'$ and 
$\sin 2\alpha'$, while allowing $\rho$, $\eta$, $A$ and $m_t$ to vary.
We display our results in Table 2, where analyses I and II correspond
to two different choices for the inputs $\sin 2\beta'$ and $\sin
2\alpha'$.

In both analyses we first do the $\chi^2$ minimization without any input
for $\sin 2\beta'$ and $\sin 2\alpha'$, to obtain central values and 
errors on $\rho$ and $\eta$ (these would correspond to the allowed
regions of Figs. 1 without including the constraints from the straight
lines representing $Arg(M_{12}^{B_d}$)).
In analysis I, we then include as inputs,
$\sin 2\beta'$ and $\sin 2\alpha'$ calculated using these central
values, and repeat the $\chi^2$ minimization 
to obtain a new minimum $\chi^2$ and central values for $\rho$ and
$\eta$. These are the values displayed
in Table 2. Analysis II follows the same procedure, except that 
the inputs $\sin 2\beta'$ and $\sin 2\alpha'$ are calculated
using values of $\rho$ and $\eta$ that are one standard deviation
above the central values obtained in the first part of the procedure
(the central values and errors obtained here would correspond to the
allowed regions of Figs. 1 where we hsave included all the constraints
including those from $Arg(M_{12}^{B_d}$)). 
In both the analyses we include an experimental error on 
$\sin 2\beta'$ of $\pm 0.059$ and on $\sin 2\alpha'$ of $\pm 0.085$
which are the errors quoted by the BABAR colloboration in
\cite{Babar}.{\footnote{Although the determination of $\sin 2\beta'$ is
not affected by ``penguin pollution'' in this model, the determination
of $\sin 2\alpha'$ could be affected by out-of-phase supersymmetric
penguins in addition to the standard model ones. We assume that these
effects could be accounted for by an isospin analysis 
\cite{Gronau}.}}
 
The three cases in Table 2 correspond to those of
Fig. 1 {\it i.e.} case (a) corresponds to the standard model where
there is no new physics, in case (b) we correctly include the
supersymmetric contribution, whereas in case (c) we include the
supersymmetric contributions but neglect the out-of-phase part.
We have checked that the results of our $\chi^2$ analysis for the
standard model agree with those of \cite{London2} for similar choices of
inputs. 

\begin{table}
\begin{center}
\begin{tabular}{|ll|l|l|}
\hline
 & Analysis & $(\rho,\eta)$ & $\chi^2_{\rm min}$ \\
\hline
I & a) Standard Model & $(-0.04\pm 0.03,0.35\pm 0.04)$ & 0.013 \\
  & b) Correct Susy   & $(0.12\pm 0.03,0.34\pm 0.03)$ & 0.0051\\
  & c) Incorrect Susy & $(0.13 \pm 0.03,0.36 \pm 0.03)$ & 0.88 \\
\hline
II& a) Standard Model & $(0.17\pm 0.04,0.42\pm 0.03)$ & 2.0 \\
  & b) Correct Susy   & $(0.23\pm 0.05,0.39\pm 0.03)$ & 1.9\\
  & c) Incorrect Susy & $(0.25\pm 0.05,0.42\pm 0.03)$ & 3.8\\
\hline
\end{tabular}
\end{center}
\caption {Results of the $\chi^2$ analysis for $\rho$ and
$\eta$ based on the inputs of Table 1.}
\end{table}

We see that the results presented in Table 2 corroborate the visual
information of Fig. 1. Firstly the central values for $\rho$ and
$\eta$ for the supersymmetric case are indeed different from those for
the standard model. In particular more positive values for $\rho$ are
preferred by the supersymmetric case. Secondly, it is only if the
actual values for $\rho$ and $\eta$ were to lie near their current $1
\sigma$ upper bounds, as in analysis II, 
that the $B$ factory experiments would be
sensitive to the new phase in $M_{12}^{B_d}$. This is signalled by the
large value of the minimum $\chi^2$ for the incorrect analysis in II
where we assume that there are no new phases in $M_{12}^{B_d}$
(since we have seven experimental inputs and four variables, we consider 
$\chi^2 < 3$ indicative of a good fit). 
This is as in Fig. 1(c) where 
we can see that for $\rho$ and $\eta$ near their central
values, the area predicted by the incorrect analysis would lie within
the allowed region, whereas with $\rho$ and $\eta$ close to
their $1\sigma$ upper bounds, the area predicted by the incorrect
analysis clearly lies outside the allowed region. Thus, it seems that
even with the precise phase information provided by the $B$ factory
experiments, we would still have to be lucky in order to be able to
notice any deviation from the usual expectations of no new phases in
neutral meson mixing. However,
the insensitivity of this analysis to the new phase is
mostly due to the large errors in the experimental and theoretical
inputs into the analysis. Since we expect most of these to decrease
before the $B$ factory data analyses begin, we redo the analysis of
Table 2 using the same central values for the inputs, but with the
improved errors expected in the future. 

We display our new inputs in Table 3, and the results of our $\chi^2$
analysis in Table 4. We base our estimates for the improved errors on
the inputs on ~\cite{Buras}.  Figs. 2 display the
same constraints as Figs. 1, 
but are plotted using the reduced errors of Table 3. 

\begin{table}
\begin{center}
\begin{tabular}{|l|l|}   
\hline
Parameter & Value \\
\hline
$|V_{ub}|/|V_{cb}|$ & $0.08 \pm 0.01$ \\
$\Delta m_{B_d}$ &  $(0.306 \pm 0.0158) \times 10^{-12}$ GeV \\
$|\epsilon|$ & $ (2.26 \pm 0.02) \times 10^{-3}$ \\
$|V_{cb}|$ & $0.039 \pm 0.001$ \\
$\lambda$ & 0.2205 \\
$m_t$ & $170 \pm 5$ GeV \\
$m_c$ & $1.3$ GeV \\
$\sqrt{B_B}f_B$  & $180 \pm 10 MeV$ \\
$B_K$ & $0.8 \pm 0.05$ \\
$\eta_B$ & 0.55 \\
$\eta_K, ~\eta_{ct}, ~\eta_{cc}$ & 0.57, 0.47, 1.32 \\
$\tilde m, ~\tan\beta$ & 85 GeV, 1 \\ 
\hline
\end{tabular}
\caption {Inputs for the $\rho-\eta$ analysis presented in
Fig. 2 and Table 4. The central values are the same as those of Table
1, however the errors reflect our expectations for experimental and
theoretical improvements in estimating these quantities. }
\end{center}
\end{table}
\begin{table}
\begin{center}
\begin{tabular}{|ll|l|l|}
\hline
 & Analysis & $(\rho,\eta)$ & $\chi^2_{\rm min}$ \\
\hline
I & a) Standard Model & $(-0.05\pm 0.03,0.35\pm 0.02)$ & 0.07 \\
  & b) Correct Susy   & $(0.12\pm 0.02,0.34\pm 0.02)$ & 0.03\\
  & c) Incorrect Susy & $(0.12 \pm 0.02,0.34 \pm 0.02)$ & 4.1 \\
\hline
II& a) Standard Model & $(0.03\pm 0.02,0.37\pm 0.02)$ & 1.1 \\
  & b) Correct Susy   & $(0.16\pm 0.02,0.35\pm 0.02)$ & 1.1\\
  & c) Incorrect Susy & $(0.16\pm 0.03,0.36\pm 0.02)$ & 6.0\\
\hline
\end{tabular}
\caption {Results of the $\chi^2$ analysis for $\rho$ and
$\eta$ based on the inputs of Table 3.}
\end{center}
\end{table}

\begin{figure}
\epsfig{file=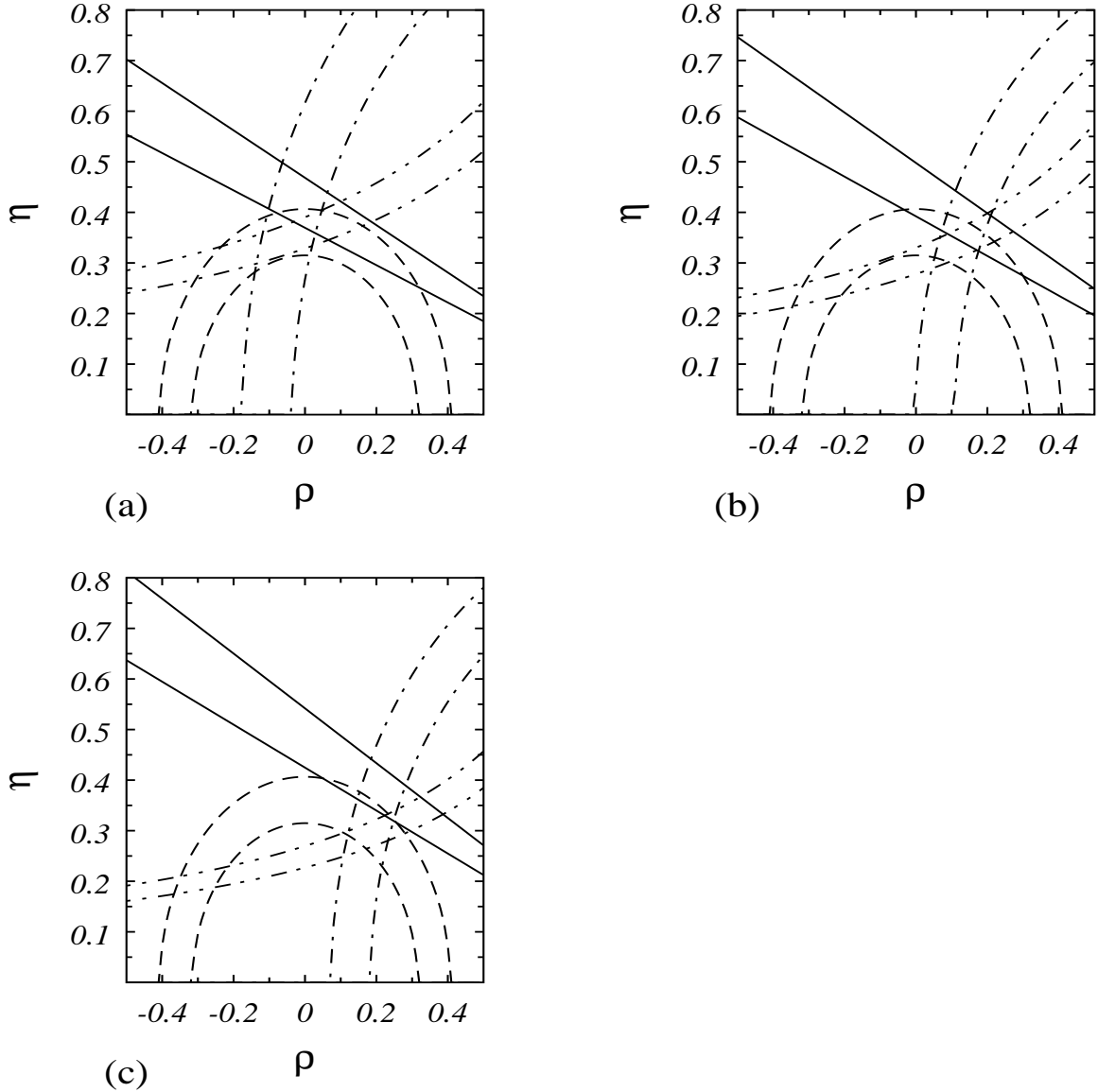,height=6in,width=6in}
\caption {Constraints on $\rho$ and $\eta$ based on the
reduced errors on the input parameters we expect in the future (Table
3). The small circles (dash) are from
$|V_{ub}|/|V_{cb}|$, the large circles (dot dash) are from
$\Delta m_{B_d}$, the hyperbolae (dot dot dash) from $\epsilon$ and the
straight lines (solid) from $Arg(M_{12}^{B_d})$.
(a) The standard model. (b) The correct supersymmetric
analysis. (c) The incorrect supersymmetric analysis.}
\end{figure}

Table 4 (as well as Fig. 2) contains what we believe to be an accurate
representation of
the physics results obtained at the $B$ factories if the scenario
outlined in this paper were to hold, {\it i.e.}, the existence of low
energy supersymmetry with small $\tan\beta$, light right-handed
up-type squarks and no new CP violating phases. Once again we notice
that the central value for $\rho$ is more positive and clearly
different from what would be the standard model value. Here however,
in contrast with the results of Table 2, we see that incorrectly
assuming that $Arg(M_{12}^{B_d})$ is not affected by the new physics
yields a poor fit over most of the allowed region for $\rho$ and
$\eta$ (cases I(c) and II(c)). 
Interestingly though, this deviation from the standard model
is not due to the existence of new $CP$ violating phases in the
theory as one would naively infer, but simply due to the fact that
the mixing patterns of the
squarks could be different from those of the quarks, resulting in the
CKM phase showing up in physical quantities in combinations different
from those in the standard model. It is exciting to know that the
experiments at the proposed $B$ factories are sensitive to this
possibility. Although we have based our analysis on one particular
choice for $\tan\beta$ and the mass of the lightest squark, the
explicit formulas presented make generalizations to other values
trivial. In particular, the effects we discuss become larger for
smaller squark mass and $\tan\beta$, and are reduced in the opposite
limit (as long as $\tan\beta \la 30$, after which this analysis no
longer holds). 

\section{Conclusions}

We have presented a scenario based on the MSSM where although the CKM
paradigm for $CP$ violation still holds, substantial
mixing between the right-handed top and charm squarks introduces new
$CP$ violating phases into the neutral meson mixing matrices.   
We have analyzed a specific case where the lightest right-handed
squark weighs $85$ GeV and $\tan\beta=1$, and show that in this case, the
experiments at the proposed $B$ factories would be sensitive to these
new phases. We stress that the presence of these new phases 
in the meson mixing matrices should not
be interpreted as proving the existence of new fundamental $CP$
violating phases, but rather as a novel manifestation of the 
usual CKM phase due to
different patterns of mixing for the quarks and the
squarks. 

\bigskip

{\bf {\large Acknowledgements}}
\medskip

I would like to thank M. Beneke, A. Grant, D. London, J. Rosner,
H. Quinn and J. Wells for useful discussions.

%\newpage


\begin{thebibliography}{99}
\bibitem{DefMssm}
{By Minimal Supersymmetric Standard Model we mean the supersymmetric
extension of the standard model with minimal particle content. Its
constrained version includes the assumptions of coupling constant
unification, and universal scalar and gaugino masses at a high
``unification'' scale.}
\end{thebibliography}
\end{document}